\documentclass{article}
\usepackage{graphicx}

\begin{document}

\title{On the possibility of extending the Nore-Frenkel generalized law of
  correspondent states to non-isotropic patchy interactions.}
\author{Giuseppe~Foffi$^*$ and Francesco~Sciortino$^\dagger$ \\ \\$^*$
  Institut Romand de Recherche Num\'erique en Physique \\ des Mat\'eriaux (IRRMA)
  and Institute of Theoretical Physics (ITP), \\Ecole Polytechnique
  F\'ed\'erale
  de Lausanne (EPFL),\\ CH-1015 Lausanne, Switzerland \\ \\
  $^\dagger$Dipartimento di Fisica and INFM-CNR-CRS Soft,\\ Universit\`a di Roma
  {\em La Sapienza}, P.le A. Moro 2, 00185 Roma, Italy}

\maketitle
 
\begin{abstract}
Colloidal systems (and protein solutions) are often characterized by
attractive interactions whose range is much smaller than the particle
size.  When this is the case and the interaction is spherical, systems
obey a generalized law of correspondent states (GLCS), first proposed
by Noro and Frenkel [~J.Chem.Phys. 113, 2941 (2000)~].  The
thermodynamic properties become insensitive to the details of the
potential, depending only on the value of the second virial
coefficient $B_2$ and the density $\rho$.  The GLCS does not
generically hold for the case of non-spherical potentials.  In this
Letter we suggest that when particles interact via short-ranged
small-angular amplitude patchy interactions (so that the condition of
only one bond per patch is fulfilled) it is still possible to
generalize the GLCS close to the liquid-gas critical point. 
\end{abstract}

\maketitle

\newpage

In colloidal and in protein systems, the interaction potential is often
short-ranged, i.e. small as compared to the particle size.  The small-range
interaction brings some peculiar behaviour both on the system thermodynamics
and on the dynamics. For example, the gas-liquid phase separation becomes
metastable with respect to crystallization~\cite{Gast1983} (and hence a proper
equilibrium liquid phase is missing).  In addition, for very short ranges, the
kinetic arrest (glass) line becomes reentrant and two different glass phases
appear~\cite{Sciortino2002,Dawson2001}.  The presence of short-range
attractions is also invoked to explain the so-called ``crystallization slot''
in the phase diagram of globular proteins~\cite{Wolde1997}.

A common property unifies all spherically symmetric short ranged
attractive potentials, independently from their actual  shape.
Indeed, Noro and Frenkel~\cite{Noro2000} showed that the
thermodynamical properties of systems interacting with short-range
attractive potentials are all equivalent if scaled by the proper
variables, i.e. they obey a generalized law of correspondent stated
(GLCS). They showed that the virial coefficient can be used as scaling
variable for the strength of the interaction.  Recently, the GLCS has
been shown to arise from the fact that, due to the short-range of the
interaction, each interacting pair of particles (a bond) contributes
independently and equally to the partition function~\cite{Foffi2006}.

The above considerations are valid for centrosymmetric potentials and can not
be straightforwardly extended to non-spherical
cases~\cite{Kern2003,Charbonneau2007}, i.e. when interactions are patchy and
strongly directional.  In the case of molecular fluids, patchy interactions
are relevant in network-forming systems, like silica~\cite{Vega1998,De2006}
and water~\cite{Kolafa1987a,De2006a} and in all associating
fluids~\cite{Chapman1988,Sear1996} in which the hydrogen bond plays an
important role. For colloidal systems, this interest is justified by recent
advances in the synthesis and characterization of patchy
particles~\cite{Manoharan2003,Cho2005,Zerrouki2006,Zhang2005} or
functionalized colloidal particles~\cite{Mirkin1996}. For what concern
proteins, it is well established that the interactions are intrinsically
directional~\cite{Sear1999,Lomakin1999,Liu} and that their description in term
of isotropic potentials represents only a first-order approximation.  \\
Recent progress in understanding the thermodynamic of patchy colloidal
particles has been based on application of the thermodynamical perturbation
theory developed by Wertheim~\cite{Wertheim1984a}. The theory, which does not
account for the geometry of the patches, assumes that each patch acts as an
independent interacting unit. In addition, the theory neglects the possibility
of close-loops of bonds.  Despite these approximations, several predictions of
the theory have been numerically
confirmed~\cite{Chapman1988,Bianchi2006,Sciortino2007}.  It has been shown
that (for the case of patchy interactions) the number of {\it independent}
interacting patches (the valence) is the key ingredient in controlling the
phase diagram of the system. On lowering the valence, the liquid-gas critical
point shifts to smaller and smaller densities, so that liquid states of
vanishing density (empty liquids) become accessible~\cite{Bianchi2006}. The
Wertheim expression for the bonding free-energy is a function only of the bond
probability and of the valence. In this respect, the theory suggests that
systems with the same valence should behave similarly, if the bond probability
is the same.  Since the bond probability is related to the chemical bonding
constant~\cite{Chapman1988,Sciortino2007}, and since for large attraction
strengths the chemical constant is proportional to the second virial
coefficient~\cite{Sear1999}, the Wertheim theory suggests that universality
based on the virial scaling can be recovered also in the case of patchy
interactions when valence is preserved.

\begin{figure}[tbh]
\includegraphics[width=.7\textwidth]{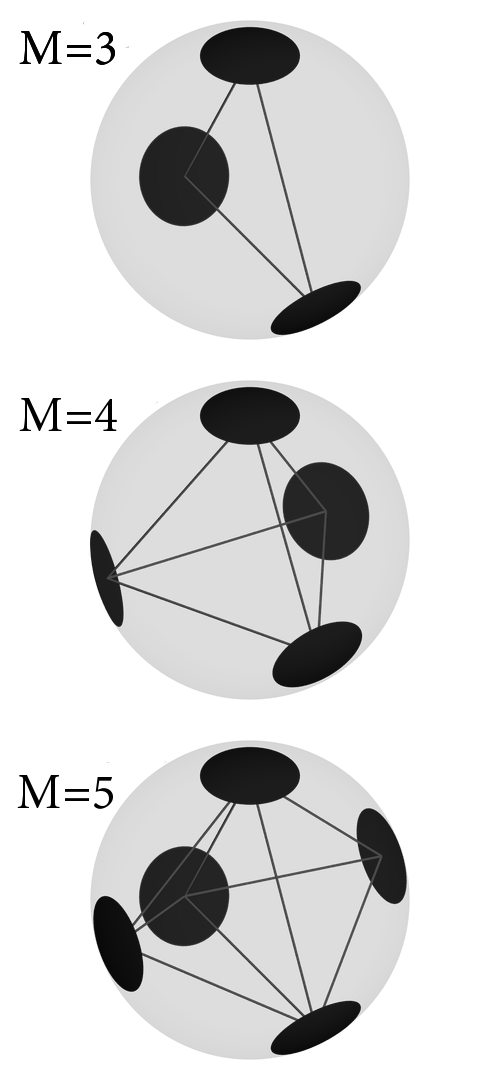}
\caption{Pictorial representation of the three different patch geometries considered in this work.
The solid angle of each patch is  $2 \pi  (1-\cos \theta)$, where $\theta$ is the cone semi-angle.
}
\label{fig1}
\end{figure}

To address the issue of a possible generalization of the Noro-Frenkel
scaling we have extended the investigation concerning the location of
the critical point for the patchy model introduced by Kern and
Frenkel~\cite{Kern2003}, for several values of the attraction range
and of the width of the patchy attractive region.  The choice of this
potential is motivated by the fact that angular and radial properties
can be independently modified.  More precisely, the two body potential
is defined as:
\begin{equation}
\label{eq1}
u(\mathbf{r}_{ij})=u^{sw}(r_{ij})f(\{\Omega_{ij}\})
\end{equation}
where $ u^{sw}(r_{ij})$ is an isotropic square well term of
depth $u_0$ and attractive range $\sigma+\Delta$ and
$f(\{\Omega_{ij}\})$ is a function that depends on the orientation of
the two interacting particles $\{\Omega_{ij}\}$. The diameter of the
particles and the depth of the square well has been chosen as units of
length and energy respectively, i.e.  $\sigma=1$ and $u_0=1$. Each
particle is characterized by $M$ identical patches. A patch $\alpha$
is defined as the intersection of the surface of the sphere with a
cone with half-opening angle $\theta$ that has the vertex in the
center of the particle and it has the axis directed toward the
direction ${\bf \hat{u}_\alpha}$. The angular function
$f(\{\Omega_{ij}\}) $ is defined as
\begin{equation}
\label{eq2}
f(\{\Omega_{ij}\}) = \left\{
  \begin{array}{ll}
1 &\mbox{if \, }\left\{
  \begin{array}{lll} \mbox{ $\hat{\bf r}_{ij}\cdot\hat{\bf u}_{\alpha}>\cos
      \theta$ } & \begin{array}{l} \mbox{some patch $\alpha$} \\\mbox{
        on particle $i$}\end{array}\\\mbox{and}&\\ 
    \mbox{ $\hat{\bf r}_{ij}\cdot\hat{\bf u}_{\beta}>\cos
      \theta$ } & \begin{array}{l} \mbox{some patch $\beta$} \\\mbox{
        on particle $j$}\end{array}
            \end{array} \right.\\
0 & \mbox{else}
\end{array} \right.
\end{equation}
where $\hat{\bf r}_{ij}$ is the direction of the vector that joins the
centers of the two interacting particles and $\alpha\,(\beta)$ some
patch belonging to the particle $i\, (j)$.
In practice two particles interact attractively if, when they are within the
attractive distance $\sigma+\Delta$, two patches are properly facing each
other. When this is the case, the two particles are considered bonded.
Decreasing $\Delta$ reduces the range of the attraction whereas reducing $
\theta$ diminishes the angular size of the patches. In the limit $\Delta
\rightarrow 0$ the model goes toward the patchy Baxter limit. In the limit
$\cos \theta \rightarrow 1$, the patch goes to the point limit.\\ In the
present work we focus on M=3, M=4 and M=5 patches, located on the surface of
the particle as shown in the cartoon of Fig.~\ref{fig1}. Differently from
previous studies~\cite{Kern2003}, we consider values of $\Delta$ and $\theta$
such that, due to steric reasons, each patch is involved simultaneously in
only one pair interaction, i.e. those values fullfilling the condition $ \sin
\theta > {\frac{\sigma^2}{2(\sigma+\Delta)^2}}$. Under this single-bond per
patch condition, the number of patches coincides also with the maximum number
of possible bond per particle.

The Kern-Frenkel potential posses an analytical expression for the second
virial coefficient:
\begin{equation}
\label{eq3}
\frac{B_2}{B_2^{{\tiny HS}}}=1-\chi^2((1+\Delta)^3-1) (e^{1/T}-1)
\end{equation}
where $\chi = M \frac{1-\cos (\theta)}{2}$ is the percentage of surface
covered by the attractive patches and the temperature is measured in reduced
units, i.e. $k_B=1$.  Here $B_2^{{\tiny HS}}$ is the hard-sphere component of
the virial coefficient.  To calculate the location of the gas-liquid critical
point we perform grand canonical Monte Carlo (GCMC)
simulations~\cite{Frenkel2001}, complemented with histogram reweighting
techniques to match the distribution of the order parameter $\rho - s e$ with
the known functional dependence expected at the Ising universality class
critical point~\cite{Wilding1997}. Here $e$ is the potential energy density,
$\rho$ the number density and $s$ is the mixing field parameter. We did not
performed a finite size study, since we are only interested in the trends with
the range $\Delta$ and the angular size of the patches $\theta$.  We have
studied systems of size $L=6$ for $M=4$ and $5$ and $L=7$ for $M=3$, 
  where $L$ is the  side length of the cubic simulation box.  For each
studied $M$ --- using the methods described in ~\cite{Romano2007} --- we
calculated the critical temperature $T_c$ and density $\rho_c$ for values of
$\Delta$ between $0.119$ and $0.01$ (at fixed $\cos \theta=0.92$) and for
$\cos \theta$ between $0.895$ and $0.99$ (at fixed $\Delta=0.119$). The results
are summarized in Table~\ref{table}.

\begin{figure}[tbh]
\includegraphics[width=.9\textwidth]{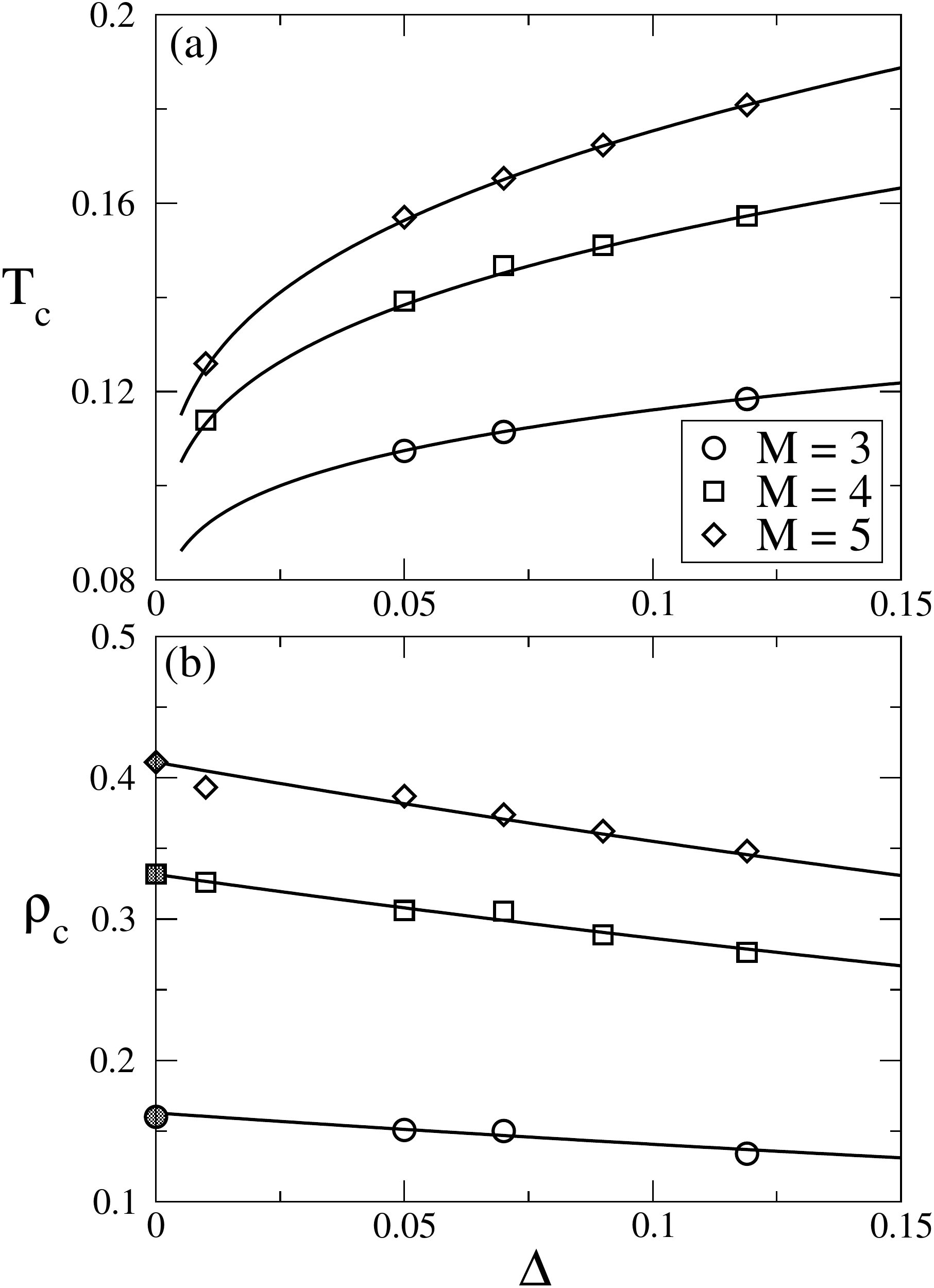}
\caption{ Critical temperature $T_c$ (a) and critical density $\rho_c$ (b) as
  a function of the range $\Delta$, at fixed patch angular size $\cos
  \theta=0.92$.  Lines in (a) correspond to constant values of $B_2$:
  specifically $B_2/B_2^{HS}=-3.03$ for $M=5$, $B_2/B_2^{HS}=-4.92
  $ for $M=4$ and $B_2/B_2^{HS}=-26.71 $ for $M=3$.  Lines in (b) are
  $\rho_c(\Delta)=\rho_c(0)/(1+0.5 \Delta)^3$, where $\rho_c(0)$ (shaded
  point) is a fitting parameter.  The fit values are $\rho_c(0)=0.41$ for
  $M=5$, $\rho_c(0)=0.33 $ for $M=4$ and $\rho_c(0)=0.16 $ for $M=3$. Open
  symbols are simulation results. }
\label{fig2}
\end{figure}

We start by analyzing the results as a function of $\Delta$, at fixed $\cos
\theta=0.92$. Fig.~\ref{fig2} shows that $T_c$ decreases with $\Delta$ while
$\rho_c$ increases.  The $\Delta$ dependence of $\rho_c$ can be conveniently
described by $\rho_c(\Delta)=\rho_c(0)/(1+0.5 \Delta)^3$, a functional form
which suggests that $\rho_c$ would be constant if measured using as unit of
length the average distance between two bonded particles ($1+0.5 \Delta$).
The resulting $\rho_c(0)$ extrapolated value provides an estimate of the
corresponding patchy Baxter model $\rho_c$. Fig.~\ref{fig2}-(a) shows also
that the $T_c$ dependences are apparently well described by iso-$B_2$ lines.
Each $M$ is characterized by a different $B_2$ value, enforcing the existence
of a GLCS for each valence. As a confirmation, we evaluate the values of
  $B_2/B_2^{HS}$ at the critical point for the patchy particle model studied
  in Ref.~\cite{Bianchi2006}. The resulting values for $M=3$,$4$ and $5$ are
  respectively $B_2/B_2^{HS} = -28.12$, $-4.95$ and $-2.78$, very similar to
  the values reported in the present study. We also note that the $B_2$
  differences between different M values are much larger than the variation
  with $\Delta$ at constant M.  Hence at a zero-th order approximation, when
  the repulsive part of the potential is complemented by localized patchy
  interactions, B2 can be considered as a scaling variable of a GLCS in the
  single-bond per patch condition.\\
  A closer look to the actual $B_2$ values (see
Table~\ref{table}) shows that a small trend in the $B_2$ values is
present, which is hidden in the logarithmic transformation relating
$B_2$ to $T$ (see Eq.~\ref{eq3}).  Still, the $B_2$ differences
between different $M$ values are much larger than the variation with
$\Delta$ at constant $M$.  This suggests that, at a zero-th order
approximation, $B_2$ can be considered as a scaling variable of a GLCS
in the single-bond per patch condition.  Thus, provided that the
geometry of the patches is such that their number coincides with the
maximum valence, $B_2$ carries the information on the valence of the
patchy interaction potential.  These results suggest that,
statistically, configurations with the same Boltzmann weight are
generated under an isotropic scaling (to change the inter-particle
distances preserving the same bonding pattern) and a simultaneous
change of both $\Delta$ and $T$ such that $B_2$ remains constant. It
is interesting to observe that the GLCS for the isotropic square well
is fulfilled for values of the range smaller then
$0.05$~\cite{Foffi2006,Malijevsky2006}, whereas for the three models
discussed here GLCS appears to hold for longer ranges.

\begin{figure}[tbh]
\includegraphics[width=.9\textwidth]{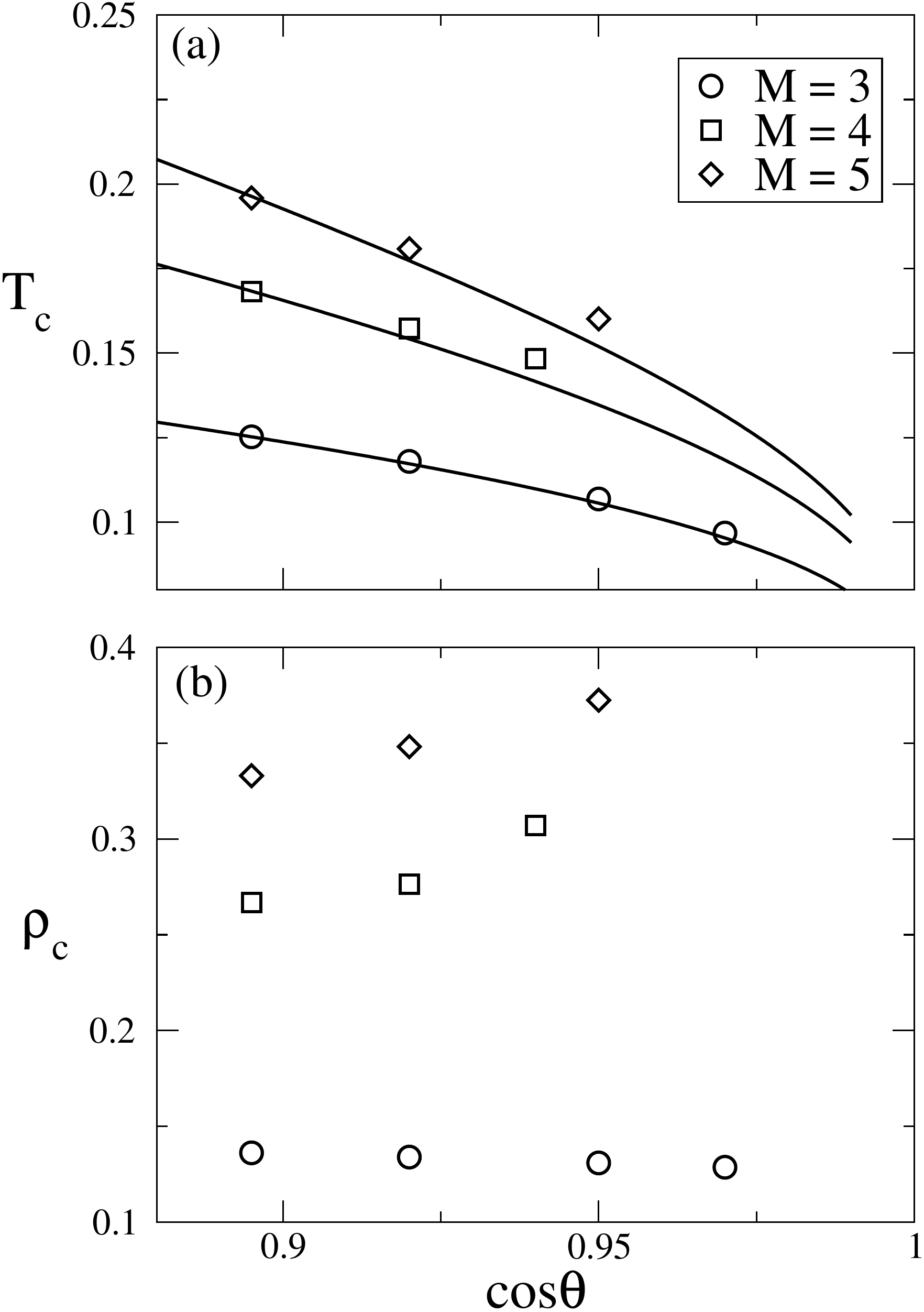}
\caption{Critical temperature $T_c$ (a) and critical density $\rho_c$ (b) as a
  function of the angular patch size $\cos \theta$, at fixed range
  $\Delta=0.119$.  Lines in (a) correspond to constant values of $B_2$:
  specifically $B_2/B_2^{HS}=-3.53$ for $M=5$, $B_2/B_2^{HS}=-5.74
  $ for $M=4$ and $B_2/B_2^{HS}=-28.29 $ for $M=3$. Open symbols are
  simulation results.  }
\label{fig3}
\end{figure}

Fig.~\ref{fig3} shows $T_c$ and $\rho_c$, this time at $\Delta=0.119$, as a
function of $\cos \theta$. For values of $\cos \theta$ larger than the one
reported in the figure, crystallization is observed within the simulation
time, preventing the possibility of evaluating the critical parameters.  This
effect suggests that the liquid-gas critical point is metastable, as in the
case of spherical short-range potentials.  The observation of crystallization
inform us already that on reducing $\theta$, some bonding patterns acquires a
larger statistical weight.  Indeed, differently from the previous case, small
deviations from a GLCS are observed in $T$ (see lines Fig.~\ref{fig3}-(a)) and
not only in $B_2^c$. We attribute these deviation to the fact that by changing
the angular part of the potential, the statistical relevance of specific
bonding patterns varies.  In other words, it is not possible to vary the
orientation of the particles to preserve the bonding on changing $\theta$,
i.e. it is not possible to perform the operation equivalent to rescaling the
distances to preserve the bonding pattern on changing $\Delta$.  We expect
that the breaking of the scaling will be enhanced at state points far from the
critical region where an extensive bonding pattern is present, i.e.  low $T$
and large $\rho$.  Concerning the $\theta$ dependence of $\rho_c$, we note
that it increases with $\cos \theta$, for the case $M=5$ and $M=4$ while
it weakly decreases for $M=3$. We have no clear arguments for interpreting the
$\rho_c(\theta)$ trends, except for the fact that the observed dependence
(more significant for $M=4$ and $M=5$), suggests a non-trivial coupling between
the angular correlation induced by bonding and the density.  For these two $M$
values, small bonding angles appear to require smaller average distances
between next-nearest neighbor particles implying larger densities.

The argument behind the quasi-validity of the GLCS for each $M$ class are
based on the fact that the relevant role is played by the bonding
pattern~\cite{Foffi2006}, which is supposed to be statistically identical for
all members of the class along corresponding states.  Hence, the number of
bonds at the critical point should be similar.  To double check this statement
we also report in Table~\ref{table} the bond probability $p_b^c$, defined as
the potential energy at the critical point normalized by the energy of the
fully bonded system.  Such quantity is indeed constant for each $M$ value, in
agreement (and strongly supporting) the possibility of defining a different
GLCS for each valence class. We also note that, within each $M$ class, the
variation of $p_b^c$ are smaller than the one of $B_2$, suggesting that the
bond probability may result in a better scaling variable than $B_2$. A small
trend in $p_b^c$ is only observed in the $\theta$ dependence.  We also note
that this observation is in agreement with the Wertheim
theory~\cite{Wertheim1984a} and with the identification of the bond free
energy as the appropriate scaling variable for the spherical
case~\cite{Foffi2006}.

In summary, we have provided evidence that different short-ranged
non-spherical potentials, but with the same number of single-bond
patches, essentially obey a GLCS.  The condition of a single-bond per
patch requires that both the attraction range and the angular size of
the patches are small.  Breakdowns of the GLCS can be expected for
potentials which differ in their angular part, especially for very
small angular sizes ($\cos \theta \rightarrow 1$), since --- in
conditions of extensive bonding --- the statistical weight of closed
loops of bonds becomes significantly affected by the angular patch
size.
\\ \\
We acknowledge support from  Swiss National Science Foundation
Grant No. 99200021-105382/1  (GF)  and MIUR PRIN (FS). 

\bibliographystyle{jpc}


\newpage

\begin{table}[h]
\begin{center}
\begin{tabular}{|c|c|c|c|c|c|c|}
\hline
$M$ & $\Delta$ & $\cos \theta$ & $T_c$ & $\rho_c$ & $p_b^c$ &
$B_2^c/B_2^{HS}$ \\
\hline \hline

3 & 0.050 & 0.92 & 0.1076 &  0.151 & 0.727 & -23.67 \\
3 & 0.070 & 0.92 & 0.1113 &  0.150 & 0.732 &-24.92 \\
3 & 0.119 & 0.92 & 0.1180 &  0.134 & 0.721 &-26.71 \\
\hline
3 & 0.119 & 0.895 & 0.1252 & 0.136  & 0.737 &-28.29 \\
3 & 0.119 & 0.92 & 0.1180 &  0.134 & 0.721 &-26.71 \\
3 & 0.119 & 0.95 & 0.1069 &  0.131 & 0.726 &-25.13 \\
3 & 0.119 & 0.97 & 0.0968 &  0.129 & 0.726 &-23.92 \\
\hline
\hline
4 & 0.010 & 0.92 & 0.1140 & 0.326 & 0.656 & -4.02 \\
4 & 0.050 & 0.92 & 0.1392 & 0.306 & 0.652 & -4.32 \\
4 & 0.070 & 0.92 & 0.1468 & 0.306 & 0.641 & -4.24 \\
4 & 0.090 & 0.92 & 0.1511 & 0.289 & 0.642 & -4.65 \\
4 & 0.119 & 0.92 & 0.1573 & 0.276 & 0.644 & -4.92 \\
\hline
4 & 0.119 & 0.895 & 0.1682 & 0.267 & 0.635 & -5.74 \\
4 & 0.119 & 0.92 & 0.1573 & 0.276 & 0.644 & -4.92 \\
4 & 0.119 & 0.94 & 0.1484 & 0.307 & 0.667 &  -3.85\\
\hline
\hline
5 & 0.010 & 0.92 & 0.1259 & 0.393 & 0.567 & -2.41\\
5 & 0.050 & 0.92 & 0.1570 & 0.387 & 0.588 & -2.67 \\
5 & 0.070 & 0.92 & 0.1653 & 0.374 & 0.585 & -2.81 \\
5 & 0.090 & 0.92 & 0.1723 & 0.362 & 0.581 & -2.90 \\
5 & 0.119 & 0.92 & 0.1808 & 0.348 & 0.577 & -3.03 \\
\hline
5 & 0.119 & 0.895 & 0.1959 & 0.333 & 0.570 & -3.53 \\
5 & 0.119 & 0.92 & 0.1808 & 0.348 & 0.577 & -3.03 \\
5 & 0.119 & 0.95 & 0.1601 & 0.372 & 0.589 & -2.22 \\
\hline
\end{tabular}
\caption{Critical point properties for all studied models, labeled by the
  values $M$, $\Delta$ and $\cos \theta$.  The different columns indicate the
  critical temperature $T_c$, critical density $\rho_c$, the bond probability
  $p_b^c$ and the reduced value of the second virial coefficient
  $B_2^c/B_2^{HS}$ at the critical point. The estimated errors for each
  of these quantities are: $\pm 0.0005 (T_c)$; $\pm 0.007 (\rho_c)$; $\pm
  0.005 (p_b^c)$.  The error in $B_2$ arises from the error in $T_c$ and
  differs for each point due to the non-linear relation between $B_2$ and $T$.
  The field-mixing parameter $s$ is always smaller than 0.08.}
\label{table}
\end{center}
\end{table}


\end{document}